\documentclass[sigconf,natbib=false]{acmart}
\usepackage[autostyle]{csquotes}
\usepackage{todonotes}

\usepackage{booktabs} 
\copyrightyear{2018} 
\acmYear{2018} 
\setcopyright{acmlicensed}
\acmConference[MSR '18]{MSR '18: 15th International Conference on Mining Software Repositories }{May 28--29, 2018}{Gothenburg, Sweden}
\acmBooktitle{MSR '18: MSR '18: 15th International Conference on Mining Software Repositories , May 28--29, 2018, Gothenburg, Sweden}
\acmPrice{15.00}
\acmDOI{10.1145/3196398.3196443}
\acmISBN{978-1-4503-5716-6/18/05}

\begin{document}
\title{Bayesian Hierarchical Modelling for Tailoring  Metric Thresholds} 

\author{Neil A. Ernst}
\orcid{0000-0001-5992-2366}
\affiliation{%
  \institution{Department of Computer Science -- University of Victoria}
}
\email{nernst@uvic.ca}

\begin{abstract}
Software is highly contextual. While there are cross-cutting `global' lessons,
individual software projects exhibit many `local' properties. This data
heterogeneity makes drawing local conclusions from global data dangerous.
A key research challenge is to construct locally accurate
prediction models that are informed by global characteristics and data volumes.
Previous work has tackled this problem using clustering and transfer learning
approaches, which identify locally similar characteristics. This paper applies a
simpler approach known as Bayesian hierarchical modeling. We show that
hierarchical modeling supports cross-project comparisons, while preserving local
context. To demonstrate the approach, we conduct a conceptual replication of an
existing study on setting software metrics thresholds. Our emerging results show
our hierarchical model reduces model prediction error compared to a global
approach by up to 50\%.
\end{abstract}

%

\maketitle

\section{Introduction}

There are two problems with doing statistical inference on software projects.
The first is that finding small \emph{true} effects is often only possible with
large datasets, yet many software projects are comparatively small. The second
problem is that software characteristics (such as object oriented metrics)
differ from one project to the next \cite{zhang13context}. Posnett
\cite{Posnett:2011} terms this the \emph{ecological inference problem}. As a
result, many software researchers, particularly in the defect prediction and
effort estimation communities, have moved to cross-project approaches, including
clustering (identifying all projects with similar characteristics) and transfer
learning (finding a function that maps a classifier from one project onto the
characteristics of another project). Both have underlying models that are hard
to interpret, and ignore domain-specific knowledge about how software is
constructed. 

This paper's new idea is to use a simpler and more understandable model for
learning from software data, Bayesian hierarchical models (also known as
multi-level models or random effects models \cite{Sorensen16})\footnote{`modeling' is used here in
the statistical sense, not the software abstraction sense}. A hierarchical model
is one where the model parameters themselves have parameters drawn from a
probability distribution. Parameters are things like mean and variance in a
Normal distribution. Hierarchical models calculate project-specific metrics,
while still taking into account the global ecological commonalities. Among other
benefits, this means that inference on projects with small datasets (level 1)
can be improved by our knowledge of the global set of data (level 2). 

Our prediction task is to find the mean Coupling Between Objects metric for an
individual project. To solve this prediction task, we compare hierarchical
modeling to two other linear regression approaches. A \emph{global} model
(approach 1) pools all file information into a single level. Any interesting
project-level variance is lost. A \emph{local} or unpooled model (approach 2)
calculates local-only regression coefficients. No global-level data is used, so
it exhibits `anterograde amnesia'---any lessons from analyzing previous projects
are forgotten. Projects with few data points will be untrustworthy (high
standard errors). Finally, a \emph{hierarchical}, or partial pooling model
(approach 3) fits a linear model per project, but regularizing the prediction
using the global data. 

We find that partial pooling is much better than the other two approaches, using
root mean squared error (RMSE). The advantage of the partial pooling approach: 
\begin{enumerate}
	\item We have an explicit prior, so it is easy to understand what is in the model and how the model is created. 
	\item Sophisticated tooling supports hierarchical models (e.g., Stan \cite{Carpenter2017}), so off
	the shelf solutions are available.
	\item Partial pooling accommodates our intuitive understanding of software development analysis, that balances the local and global data available.
\end{enumerate}

While popular in social sciences (one author has said ``multilevel modeling
deserves to be the default form of regression \cite[p.14]{mcilreath16}'', this
form of modeling is underused in the software engineering literature. In 1998
Pickard, Kitchenham and Jones \cite{PICKARD1998811} pointed out that combining
results in software engineering could leverage new advances in Bayesian
hierarchical models. It does not seem as though there has been much take up of
this call; in \S \ref{related} we highlight the one or two studies that have.
The main issue has been the computational power required for producing the
posterior probability distributions. As we show, this is now resolved with
modern inference engines and processor power. 

We investigate hierarchical models with a conceptual replication of the study of
Aniche et al. \cite{Aniche2016} (henceforth SATT). The SATT study postulated
that the \emph{architectural role} a class plays (in particular, Spring
framework Controllers, Views, Repositories) influences the expected value of
certain software metrics. They assign a different threshold for various
Chidamerer and Kemerer (CK) metric values \cite{Chidamber_1991} based on the
role a file plays. We show how thresholds can be set more accurately using
hierarchical models.


We begin by introducing some statistical background. 
We apply a hierarchical model to the SATT dataset  \cite{Aniche2016}. We show
how our model is constructed, validate it with RMSE, and discuss how such
approaches might be applied in the future. An available Jupyter notebook
demonstrates how this works in
practice\footnote{\url{https://figshare.com/s/fd34d562ce882d1ab4d2}}.
Our emerging results show:
\begin{enumerate}
	\item Hierarchical models are easy to explain and set up;
	\item An example of using a probabilistic programming language for Bayesian data analysis;
	\item Hierarchical models support analysis that takes into account the rich diversity of empirical data in software engineering;
	\item Hierarchical models outperform purely local or purely global approaches.
\end{enumerate}

\section{Background}
For a general introduction to a Bayesian approach to statistics, people can
refer to \cite{gelman14} and \cite{mcilreath16}. Bayesian inference is built on
Bayes's theorem, $P(H | D) = \frac{ P(D | H) * P(H)}{P(D)}$, where H is our
hypothesis (i.e., architectural role influences software metrics), and D is the
data we gather. Bayesian inference calculates a \emph{posterior probability
distribution} $P(H|D)$. It does so as a consequence of the assumptions: $P(D |
H)$, the \emph{likelihood}; the parameters we wish to estimate; and $P(H)$ our
\emph{prior probability} for those parameters. We must explicitly assign a prior
probability in Bayesian statistics. 
Bayesian inference is a machine that \emph{conditions} on the data to generate a
posterior, given the assumptions. This machine is often hard to construct
mathematically, particularly in the hierarchical models we introduce in this
paper. As a result, probabilistic programming techniques are needed to compute
the posterior. 


Probabilistic programming is a programming paradigm that uses an inference
engine to fit complex and multi-level models. Variables are not deterministic
but rather stochastic. Their value comes from a probability distribution. In
order to compute this probability distribution, probabilistic programming
languages use Markov Chain Monte Carlo (MCMC) sampling (specifically,
Hamiltonian Monte Carlo and the No U-Turn Sampler). 
A good survey of probabilistic programming is available at \cite{Quarashi}.
One example of a probabilistic programming language (and the one used in this
paper) is Stan \cite{Carpenter2017}, and its PyStan library (the PyMC3 package
is another variant). The Stan probabilistic program to compute a linear
regression $y \sim \mathcal{N}(\mu,\sigma)$ with $\mu$, our linear model,
represented by $\beta_1 + \beta_2 x_i$, looks like:

\begin{verbatim}
model {  y ~ normal(beta[1] + beta[2] * x, sigma); }
data {
  int<lower=0> N; 
  vector[N] x;
  vector[N] y; }
parameters {
  vector[2] beta;
  real<lower=0> sigma; } 
\end{verbatim}

where we provide input for the data section (e.g., our vector of metric values),
and Stan does the inference to find values for the parameters section that
maximize the posterior, conditioned on the data. 




\section{Methodology}
Our research question is whether a hierarchical model is more accurate than a
pooled or unpooled regression model. As a working example, we perform a
conceptual replication of the study of Aniche et al. \cite{Aniche2016} (SATT).
This study explored how metrics thresholds could be contextualized based on
architectural role. We use their data to fit three different regression models,
then validate the accuracy of these models using RMSE. Once we have a model, we
can use that to estimate threshold values (i.e., level at which to take action
on refactoring). To identify if a file violates a threshold, one looks up its
role, and then retrieves the three thresholds and compares the file's metric
against the thresholds.

We focus on a narrower analysis than SATT, restricting it to the Chidamber and
Kemerer (CK) metric Coupling Between Objects (CBO) \cite{Chidamber_1991} and the
Spring Controller role. We choose CBO since the SATT results found it highly
correlated to architectural role, but our basic approach is easily extended. We
have 120 projects in the SATT dataset, 115 of which have Controllers, and a wide
range of values for number of files. There are 56,106 files overall. Some
projects are very tiny. The total file count of the smallest is 28, and the
number of Controller files in that project (for example) is 18 (recall that
handling such small datasets is one of the strengths of hierarchical models). We
calculate the Mann-Whitney measure and Cliff's Delta measure for Controller
metrics for all projects, compared to all other roles (including no role), and
find a large effect (0.657), just like the original paper did. 

The original study used a non-parametric test to show the effect of
architectural role, but for our purposes we will need a probability distribution
to serve as the likelihood. 
We choose a lognormal, as this is empirically accurate \cite{Herraiz2012}, and the same as that
chosen in  the SATT study \cite{Aniche2016}. We begin by taking the log of a
file's CBO score (LCBO).



\subsection{Modeling the Problem}
We will use a simple linear regression model to estimate the posterior
distribution of (L)CBO values, and do this in three different ways. We follow
the model approach of Chris Fonnesbeck \cite{fonnesbeck2016}. Fig.
\ref{fig:partial-pool} on page 3 shows the overall comparison of the three
models.

\noindent\textbf{Global Pooling}
Global pooling aggregates all individual files and fits a single linear
regression model (a fixed effects model). We predict LCBO score, $y$, as a
function of whether the file is a controller (0/1), $x$: $ y_i = \alpha + \beta
x_i + \epsilon_i$, with equivalent statistical model expressed as $y \sim \mathcal{N}(\alpha
+ \beta x, \sigma)$.

What we have created is a posterior probability distribution for the unknown
parameters $\alpha, \beta, \sigma$. We are not getting point estimates, but
rather distributions. To obtain the parameters and errors, we sample from the
posterior distribution, i.e., find the mean and S.E. of the samples.



\noindent\textbf{Unpooled}
The other extreme from global pooling is to model the data per project (N=115).
Each project gets a separate regression model, and we can compare the
co-efficients for each one. The regression model is $y_i = \alpha_{i[j]} + \beta
x_i + \epsilon_i $ for projects $j$.


Here we are favoring individual project variance. Small samples will greatly
bias our project-level estimates. McIlreath \cite{mcilreath16} calls this
``anterograde amnesia" since our model machinery `forgets' what it learned from
the previous projects. 

\noindent\textbf{Partial Pooling - Varying Slope and Intercept}
The hierarchical, \emph{partial pooling} approach adjusts the local estimate
\emph{conditional} on the global estimate. The hierarchy in our hierarchical
model is files within projects.  Partial pooling gives nuance we know exists
because of local effects from architectural role \cite{Aniche2016}, but
\emph{regularizes} that by the global parameters (another name for this is
\emph{shrinkage}).   

The partial pooling model will allow the regression model's slope and intercept
to vary. If our pooled regression equation predicts $y_i$, an observation at
point $i$, to be  $y_i = \alpha + \beta x_i + \epsilon_i$, than the partial
pooling regression equation is $y_i = \alpha_{j[i]} + \beta_{j[i]} x_{i} +
\epsilon_i$. The main difference with unpooled models is that we assign
hyperpriors to our parameters. 

The model component of the probabilistic program is 
\begin{verbatim}
model {
  mu_a ~ normal(0, 100); # hyperprior
  mu_b ~ normal(0, 100); # hyperprior
  a ~ normal(mu_a, sigma_a); # prior
  b ~ normal(mu_b, sigma_b); # prior
  # likelihood:
  y ~ normal(a[project] + b[project]*x, sigma); 
}
\end{verbatim}

with parameters $a,b$, priors for $y$, having posteriors that are normal, with
priors for the mean distributed as another normal model with mean 0 and sigma
100. This is an uninformative prior (allowing for a wide range of values, and
thus dominated by the data). Choosing priors effectively is a core skill in
probabilistic programming. 

\section{Results} 
\begin{figure}[tb]
	\centering
	\includegraphics[width=.7\columnwidth]{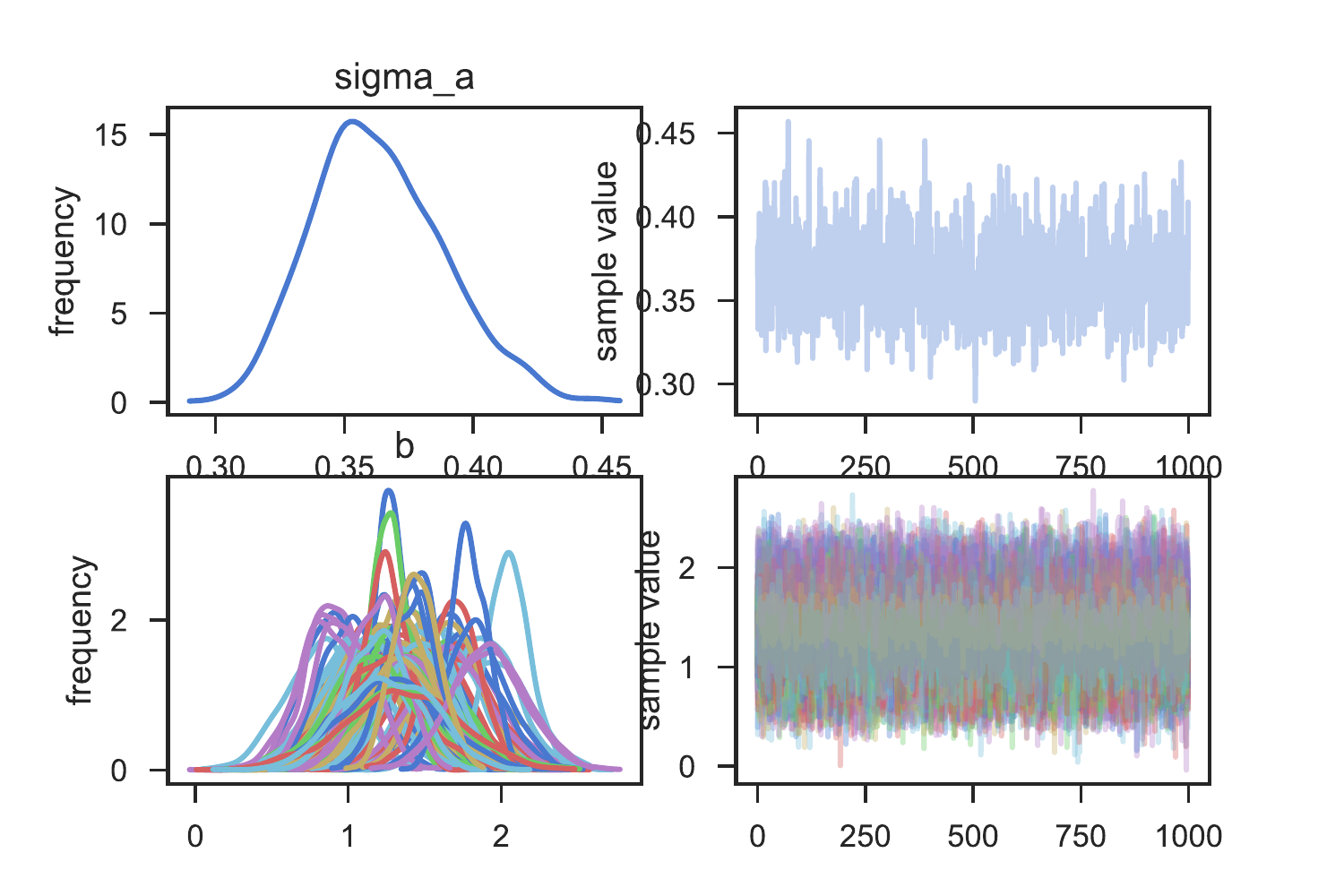}
	\caption{Checking model convergence.}
	\label{fig:modelcheck}
\end{figure}

An important part of Bayesian hierarchical modeling is model checking. This
involves inspection of the model inferences to ensure the fit is
reasonable. To do this we plot (Fig. \ref{fig:modelcheck}) the posterior distribution of our parameters and
the individual samples.

The left side shows our marginal posterior---for each parameter value on the
x-axis we get a probability on the y-axis that tells us how likely that
parameter value is. The right side shows the sampling chains. Our sampling
chains for the individual parameters (left side) seem well converged and
stationary (there are no large drifts or other odd patterns) \cite{Sorensen16}.

\begin{figure}[tb]
	\centering
	\includegraphics[width=\columnwidth]{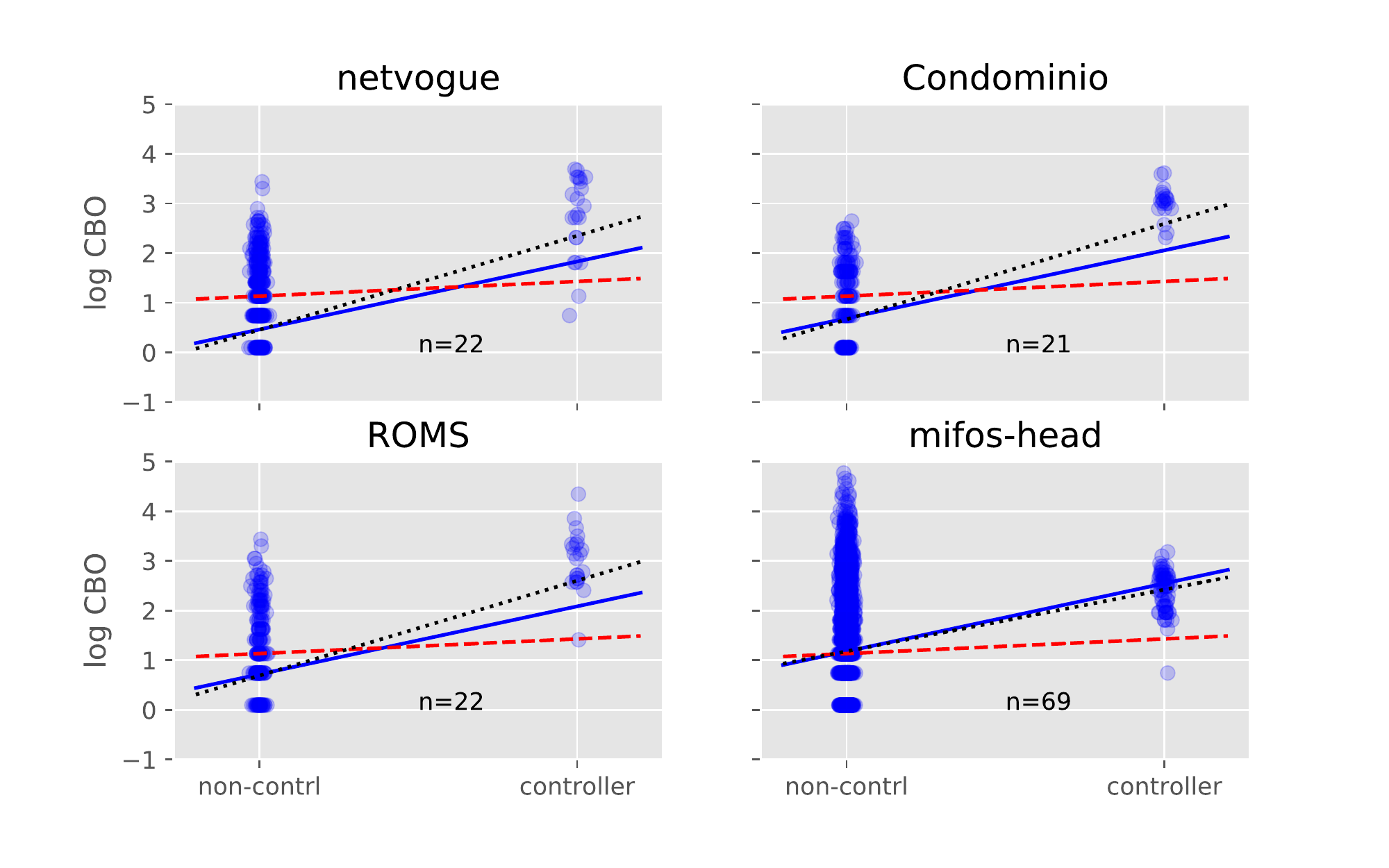}
  \caption{Comparing three models. Blue=unpooled, dotted black=partial pool, dashed red=pooled.
   N indicates number of controller files. X-axis reflects predictor of controller/not-controller.}
	\label{fig:partial-pool}
\end{figure}

A sample of four projects in the SATT dataset is shown in Fig.
\ref{fig:partial-pool}. Conceptually, the regression is more accurate if the
right side data points (in blue dots) are closer to the regression line (x=1
being files which are Controllers). The dashed black line is the partial pooling
estimate; the dashed red line is the global/pooled estimate; and the solid blue
line the unpooled estimate. The global estimate seems way off. Depending on the
number of data values, the partial and unpooled estimates are fairly close. The
improvement in accuracy of the partial pooling approach increases as the number
of datapoints decreases (and the unpooled model gets worse). 

\noindent\textbf{Validation}---We validate accuracy using a simple root mean
squared error (RMSE) approach.
We fit our three regression models, then calculate the average distance from the
predicted score to the actual file CBO values. We expect the partial-pooling to
have a lower RMSE value if it is more accurate than the others.

Table \ref{tbl:rmse} reports the mean RMSE over all projects with Controllers
(n=115), and approximate Stan model training time with 2 chains of 1000
iterations (model compilation (to C++	) is a constant \emph{30s} or so in
addition). Partial pooling is \textbf{nearly twice as effective} as full
pooling, but takes 4 times as long (an OLS approach is nearly instantaneous and
produces the same RMSE as the full pooling model). No pooling is nearly as
effective as partial pooling (likely because the number of files that are
\emph{not} controllers is fairly large). However, error values for the unpooled
approach vary; in projects with small numbers of files, the RMSE for an unpooled
approach is much higher. This represents the benefits of shrinking towards the
global mean. For example, in the Netvogue project of Fig.
\ref{fig:partial-pool}, the unpooled regression (blue) is quite far from the
partial pooling line (black dashed). This is a difference in RMSE of
1.23 vs. 0.91.

\begin{table}
\begin{tabular}{c|c|c}
\toprule
Model 			& RMSE 	& Sampling Time \\ 
\midrule
Full Pool 		& 1.099 & 26.9s \\ 
Unpooled 		& 0.630 & 5m11s \\ 
Partial Pool 	& 0.523 & 3m18s \\ 
\bottomrule
\end{tabular}
\caption{RMSE for 3 Regression Approaches}
\label{tbl:rmse}
\end{table}

\noindent\textbf{Justifying Thresholds}---The SATT finding \cite{Aniche2016} was
that a file's architectural role played a significant part in its eventual CBO
numbers. They used the procedure of Alves et al. \cite{alves2010deriving} to
identify thresholds. The Alves thresholds were set by ranking files (across all
projects) by lines of code (LOC), then finding the metric value of the file
closest to 70/80/90\% (moderate/high/very high risk). 

We use our partial pooling regression model to set thresholds at 70/80/90\% of
the normal distribution of log CBO. E.g., for the project `V2V', the partial
pooling model tells us the project's mean LCBO value $y \sim
\mathcal{N}(a_\text{project} + b_\text{project}*x, \sigma)$. This normal
posterior distribution is used to find the point probability for our thresholds
$\tau$ (i.e. the point on the normal distribution holding $\tau$\% of the
probability mass). $\text{exp}(\tau)$ produces CBO score. For example, the `V2V'
project has expected Controller CBO thresholds of 32,49, and 88. Compare this to
another project, `tatami-team', with 21,32,58, respectively. Our thresholds vary
based on the project characteristics and the regularizing global dataset. Full
results for all projects are in our replication package (footnote p. 2).

\section{Discussion and Related Work}
\label{related}
The chief threats to validity in Bayesian hierarchical modeling come from poor
understanding of the underlying probabilistic program, including what priors to
choose. Model checking is vital to mitigate this. There are subtleties in how
the sampler works that need careful checking, as shown by Wiecki's blog post
\cite{Wiecki17}. 

Other threats to validity include data analysis and implementation issues
(internal validity). One threat that is mitigated by this approach is external
validity, since a hierarchical model inherently accounts for inter-project
differences. Many of the common complaints about lack of generalizability (e.g.,
of Microsoft results to other companies) could be avoided with a hierarchical
approach.

This paper has only hinted at the uses for hierarchical modeling. Other useful
aspects include the ability to handle correlations between local and global
variables \cite{mcilreath16}, innately accounting for multiple-comparisons
problems \cite{Gelman:2012aa}, and adding group-level predictors (for example,
using number of files per project). Other aspects to explore include adding
additional predictors such as lines of code, different underlying prior
distributions, and improved analysis of model fit using Leave One Out (LOO) or
Widely Applicable Information Criterion (WAIC) \cite{vehtari15}. Although
consumer-ready, Bayesian inference remains unexplored compared to the highly
usable frequentist packages. Probabilistic programming is also an area that
merits more investigation as a form of software development in its own right.

\noindent\textbf{Related Work}---There are two main categories of \emph{related work.} 
First, there is work that identifies the \textbf{differences between per-project
and pooled predictors} in analysing software metrics. Menzies et al.
\cite{menzies11ase} introduced the concept of `local' (clusters of data with
similar characteristics) and `global' (cross-project, pooled) lessons in defect
prediction. This idea has been followed up by many others (e.g.,
\cite{Kamei2016,Bettenburg2012,panichella14}). Posnett et al.
\cite{Posnett:2011} conducted a study comparing defect prediction at two levels:
aggregated (package) vs. unaggregated (file) and introduced the concept of
ecological inference. The innovation in our paper is to propose a partial
approach, regularizing the local predictions with the global information. Our
paper does not, however, make any claim to improve on defect prediction---we
leave that to future work. Several papers \cite{Bettenburg2012,Zhou15regress}
have leveraged linear (continuous response) and logistic regression (categorical
response); however, these are single-level models. 
Bettenburg et al. \cite{Bettenburg2012} blend the regression levels using a
different approach, Multivariate Adaptive Regression Splines (MARS). The MARS
approach appears to be fairly uncommon in the statistics literature, and uses
hinge functions to reduce the residual error. The Stan models we introduce are
simpler to specify, and the Bayesian approach to conditioning on the prior data
we find more intuitive.



Jørgenson et al. \cite{JORGENSEN2003} introduced a technique called `regression
to the mean' (RTM). It also attempts to regularize local predictions using
global parameters. 
Hierarchical models are a more generalized and robust version of this, at the
expense of computation time.

There are few software engineering studies using hierarchical models. The two we
have identified are the 2012 work of Ehrlich and Cataldo
\cite{Ehrlich:2012}, who used multi-level (hierarchical) models to assess
communication in global software development. In 2017 Hassan et al.
\cite{Hassan2017} used a multi-level model to study app store reviews. This
approach is similar to this paper but uses a frequentist, maximum likelihood
approach, i.e., not Bayesian, using the `lme4' R package.
We believe the Bayesian approach better accounts for prior information that is helpful when
 lacking observations.


Secondly, there is work that looks at \textbf{metric thresholds}. Since metrics
were proposed for software, researchers have sought to identify what thresholds
should trigger warnings. The paper we replicate from Aniche et al.
\cite{Aniche2016} has a good survey. We point out the work of Herraiz et al.
\cite{Herraiz2012}, Oliveira et al. \cite{oliveira14}, Foucault et al.
\cite{Foucault2014} and Alves et al. \cite{alves2010deriving} as more recent
examples. The metrics industry, e.g., SIG and CAST, has also long used databases
of company performance (a global pool) as benchmarks for evaluating new clients.


\section{Conclusion}
In this paper we introduced a hierarchical approach to setting metric
thresholds. Using a partial pooling approach, we are able to account for global
context, while retaining local variance. Our linear regression model defined a
project-specific mean LCBO score, which we used to set the benchmarks
accordingly. The hierarchical model is simple to describe, handles cross-project
comparisons, and is more accurate than a purely global or purely local approach.

\section{Acknowledgements}
We are grateful to the authors of \cite{Aniche2016}, who made their
well-packaged and documented data and code available for others.
 
\newpage

\bibliography{/Users/nernst/Documents/refs/bib/abbr.bib,msr-satt.bib}{}

\begin{thebibliography}{10}
\providecommand{\url}[1]{#1}
\csname url@samestyle\endcsname
\providecommand{\newblock}{\relax}
\providecommand{\bibinfo}[2]{#2}
\providecommand{\BIBentrySTDinterwordspacing}{\spaceskip=0pt\relax}
\providecommand{\BIBentryALTinterwordstretchfactor}{4}
\providecommand{\BIBentryALTinterwordspacing}{\spaceskip=\fontdimen2\font plus
\BIBentryALTinterwordstretchfactor\fontdimen3\font minus
  \fontdimen4\font\relax}
\providecommand{\BIBforeignlanguage}[2]{{%
\expandafter\ifx\csname l@#1\endcsname\relax
\typeout{** WARNING: IEEEtran.bst: No hyphenation pattern has been}%
\typeout{** loaded for the language `#1'. Using the pattern for}%
\typeout{** the default language instead.}%
\else
\language=\csname l@#1\endcsname
\fi
#2}}
\providecommand{\BIBdecl}{\relax}
\BIBdecl

\bibitem{zhang13context}
F.~Zhang, A.~Mockus, Y.~Zou, F.~Khomh, and A.~E. Hassan, ``How does context
  affect the distribution of software maintainability metrics?'' in
  \emph{Proceedings of the IEEE International Conference on Software
  Maintenance}, 2013, pp. 350--359.

\bibitem{Posnett:2011}
D.~Posnett, V.~Filkov, and P.~Devanbu, ``Ecological inference in empirical
  software engineering,'' in \emph{Proceedings of the IEEE/ACM International
  Conference on Automated Software Engineering}, 2011, pp. 362--371.

\bibitem{Sorensen16}
T.~Sorensen, S.~Hohenstein, and S.~Vasishth, ``Bayesian linear mixed models
  using stan: A tutorial for psychologists, linguists, and cognitive
  scientists,'' \emph{The Quantitative Methods for Psychology}, vol.~12, no.~3,
  pp. 175--200, 2016.

\bibitem{Carpenter2017}
B.~Carpenter, A.~Gelman, M.~D. Hoffman, D.~Lee, B.~Goodrich, M.~Betancourt,
  M.~Brubaker, J.~Guo, P.~Li, and A.~Riddell, ``Stan: A probabilistic
  programming language,'' \emph{Journal of Statistical Software}, vol.~76,
  no.~1, 2017.

\bibitem{mcilreath16}
R.~McIlreath, \emph{Statistical Rethinking: A Bayesian Course with Examples in
  R and Stan}.\hskip 1em plus 0.5em minus 0.4em\relax CRC Press, 2016.

\bibitem{PICKARD1998811}
L.~M. Pickard, B.~A. Kitchenham, and P.~W. Jones, ``Combining empirical results
  in software engineering,'' \emph{Information and Software Technology},
  vol.~40, no.~14, pp. 811--821, 1998.

\bibitem{Aniche2016}
M.~Aniche, C.~Treude, A.~Zaidman, A.~van Deursen, and M.~A. Gerosa, ``{SATT}:
  Tailoring code metric thresholds for different software architectures,'' in
  \emph{Proceedings of the IEEE International Working Conference on Source Code
  Analysis and Manipulation}, oct 2016.

\bibitem{Chidamber_1991}
S.~R. Chidamber and C.~F. Kemerer, ``Towards a metrics suite for object
  oriented design,'' \emph{Proceedings of the ACM SIGPLAN Conference on
  Object-Oriented Programming, Systems, Languages, and Applications}, 1991.

\bibitem{gelman14}
A.~Gelman, J.~B. Carlin, H.~S. Stern, D.~B. Dunson, A.~Vehtari, and D.~B.
  Rubin, \emph{Bayesian Data Analysis}, 3rd~ed.\hskip 1em plus 0.5em minus
  0.4em\relax CRC Press, 2014.

\bibitem{Quarashi}
\BIBentryALTinterwordspacing
M.~AlQuraishi. (2015) The state of probabilistic programming. [Online].
  Available:
  \url{https://moalquraishi.wordpress.com/2015/03/29/the-state-of-probabilistic-programming/}
\BIBentrySTDinterwordspacing

\bibitem{Herraiz2012}
I.~Herraiz, D.~Rodriguez, and R.~Harrison, ``On the statistical distribution of
  object-oriented system properties,'' in \emph{International Workshop on
  Emerging Trends in Software Metrics ({WETSoM})}.\hskip 1em plus 0.5em minus
  0.4em\relax {IEEE}, jun 2012.

\bibitem{fonnesbeck2016}
\BIBentryALTinterwordspacing
C.~Fonnesbeck. (2016) A primer on bayesian multilevel modeling using pystan.
  [Online]. Available:
  \url{https://github.com/fonnesbeck/stan_workshop_2016/blob/master/notebooks/Multilevel%20Modeling.ipynb}
\BIBentrySTDinterwordspacing

\bibitem{alves2010deriving}
T.~L. Alves, C.~Ypma, and J.~Visser, ``Deriving metric thresholds from
  benchmark data,'' in \emph{Proceedings of the IEEE International Conference
  on Software Maintenance}, 2010, pp. 1--10.

\bibitem{Wiecki17}
\BIBentryALTinterwordspacing
T.~Wiecki. (2017) Why hierarchical models are awesome, tricky, and {B}ayesian.
  [Online]. Available:
  \url{http://twiecki.github.io/blog/2017/02/08/bayesian-hierchical-non-centered/}
\BIBentrySTDinterwordspacing

\bibitem{Gelman:2012aa}
A.~Gelman, J.~Hill, and M.~Yajima, ``Why we (usually) don't have to worry about
  multiple comparisons,'' \emph{Journal of Research on Educational
  Effectiveness}, vol.~5, pp. 189--211, 2012.

\bibitem{vehtari15}
A.~Vehtari, A.~Gelman, and J.~Gabry, ``Practical bayesian model evaluation
  using leave-one-out cross-validation and {WAIC},'' \emph{Statistics and
  Computing}, vol.~27, pp. 1413--1432, 2017.

\bibitem{menzies11ase}
T.~Menzies, A.~Butcher, A.~Marcus, T.~Zimmermann, and D.~Cok, ``Local vs.
  global models for effort estimation and defect prediction,'' in
  \emph{Proceedings of the IEEE/ACM International Conference on Automated
  Software Engineering}, 2011, pp. 343--351.

\bibitem{Kamei2016}
Y.~Kamei, T.~Fukushima, S.~McIntosh, K.~Yamashita, N.~Ubayashi, and A.~E.
  Hassan, ``Studying just-in-time defect prediction using cross-project
  models,'' \emph{Empirical Software Engineering}, vol.~21, no.~5, pp.
  2072--2106, Oct 2016.

\bibitem{Bettenburg2012}
N.~Bettenburg, M.~Nagappan, and A.~E. Hassan, ``Think locally, act globally:
  Improving defect and effort prediction models,'' in \emph{{IEEE} Working
  Conference on Mining Software Repositories}, jun 2012.

\bibitem{panichella14}
A.~Panichella, R.~Oliveto, and A.~D. Lucia, ``Cross-project defect prediction
  models: L'union fait la force,'' in \emph{Proceedings of the European
  Conference on Software Maintenance and Reengineering}, 2014, pp. 164--173.

\bibitem{Zhou15regress}
M.~Zhou and A.~Mockus, ``Who will stay in the {FLOSS} community? modeling
  participant's initial behavior,'' \emph{IEEE Transactions on Software
  Engineering}, vol.~41, no.~1, pp. 82--99, 2015.

\bibitem{JORGENSEN2003}
M.~J{\o}rgensen, U.~Indahl, and D.~Sj{\o}berg, ``Software effort estimation by
  analogy and ``regression toward the mean'','' \emph{Journal of Systems and
  Software}, vol.~68, no.~3, pp. 253 -- 262, 2003.

\bibitem{Ehrlich:2012}
K.~Ehrlich and M.~Cataldo, ``All-for-one and one-for-all?: A multi-level
  analysis of communication patterns and individual performance in
  geographically distributed software development,'' in \emph{Proceedings of
  the ACM Conference on Computer Supported Cooperative Work}, 2012, pp.
  945--954.

\bibitem{Hassan2017}
S.~Hassan, C.~Tantithamthavorn, C.-P. Bezemer, and A.~E. Hassan, ``Studying the
  dialogue between users and developers of free apps in the google play
  store,'' \emph{Proceedings of the International Symposium on Empirical
  Software Engineering and Measurement}, Sep 2017.

\bibitem{oliveira14}
P.~Oliveira, M.~T. Valente, and F.~P. Lima, ``Extracting relative thresholds
  for source code metrics,'' in \emph{Proceedings of the European Conference on
  Software Maintenance and Reengineering}, 2014, pp. 254--263.

\bibitem{Foucault2014}
M.~Foucault, M.~Palyart, J.-R. Falleri, and X.~Blanc, ``Computing contextual
  metric thresholds,'' in \emph{Proceedings of the ACM SIGAPP Symposium on
  Applied Computing}, 2014.

\end{thebibliography}
\bibliographystyle{IEEETran}
\end{document}